\documentstyle[preprint,aps]{revtex}
\draft
\begin{document}

\title{Theory of Second Order Optical Processes from
A Luttinger Liquid}

\author{Jun Zang, Joseph L. Birman}
\address{ Physics Department, The City College of The C.U.N.Y \\
Convent Avenue at 138th Street, New York, NY 10031}
\author{Zhao-Bin Su}
\address{Institute of Theoretical Physics, Academia Sinica\\
Beijing 100080, People's Republic of China}

\maketitle

\begin{abstract}
We develop a theory for the total optical secondary
emission from a 1D interacting electron system
modelled as a Tomonaga-Luttinger liquid. We separate the
emission into two parts which may originate in
{\em hot luminescence} (HL) and {\em Raman Scattering}
(RS) respectively when we neglect the {\em interference}
effect. We find a peak around $\Delta \omega
= v_f |q|$ in the RS part which does not come from a
structure factor peak. In general the
total emission cannot be separated
into HL and RS. However at resonance, and taking into
account the $k$ dependence of the optical matrix
element, a part of the RS is proportional to
the structure factor $S(q_1-q_2, \omega_1-\omega_2)$.

\end{abstract}
\vspace{0.5cm}

\pacs{Keywords: A.nanostructures A.semiconductors D.optical properties}

Recent advances in fabrication of semiconductors
permit the construction of systems, such as the
quantum wire, with effective geometrical dimensionality of one (1D).
This has permitted study of the transport properties
of 1D systems. The optical properties can also
be studied, so as to compare and contrast with electronic models
(Fermi Liquid or Luttinger Liquid)
for 1D electronic and optical response.
It is well known that the Tomonaga-Luttinger (TL) model \cite{tl}
is necessary to describe a pure 1D interacting
electron system \cite{1dgas}.
However, in an {\em actual\/} semiconductor
quantum wire, the scattering and finite size effects may
serve to damp out the low-energy plasmons and restore the
Fermi surface \cite{hdsarma}. It is not yet clear
whether the TL liquid
or the normal Fermi liquid theory is more appropriate to
describe properties of
semiconductor quantum wires. As a matter of
fact, both TL liquid and Fermi liquid theory have been
used to study the transport and optical properties of 1D
semiconductor electron systems \cite{hdsarma,kane,1dexamp,fest1,fest2}.

Optical methods are powerful tools for the investigation
of the collective and single particle excitations. For example,
the Fermi edge singularities (FES) \cite{mahan,fes}
in the photoemission and soft-x-ray
emission and absorption provide information about the
existence of the Fermi
surface and the low energy excitations.
The conditions for FES have been studied in both
normal Fermi liquid theory \cite{fesfl}
and TL liquid theory \cite{fest2,fest1}.
Although Raman Scattering (RS)
experiments have been carried out
in these 1D quantum wires \cite{1drs},
there are only a few theoretical studies
using normal Fermi liquid theory \cite{rsfl}
which aim to describe the RS spectra
of these systems. However, the calculation of the
RS efficiency from the collective excitations in 1D systems needs
special caution, especially in the {\em resonance} region,
as we discuss later. The objective of this paper
is (1) to give a theory of RS in a TL liquid; (2) to reveal that
generally the second order RS efficiency is {\em not} proportional
to the dynamical structure factor in the {\em resonance\/}
region contrary to a widely used approximation;
(3) to discuss the difference between RS from TL and that from FL.
We also remark on the possibility of seperating hot luminescence
(HL) and Raman scattering (RS). Since TL is an exactly solvable model, our
calculation
may serve as a special prototype for the discussion of various
aspects of RS, such as threshold behavior.

Let us first consider the general RS which is due to the
$\vec{A} \cdot \vec{P}$ interaction. Our discussion here
will follow the standard literature on the RS from
collective excitations \cite{mcw,rs70}.
If we here neglect the "final state interaction" of the core
holes, then in second order perturbation theory,
the transition rate can be written as:
\begin{eqnarray}
W(q_1,q_2;\omega_1,\omega_2)&=&\sum_{k,k'} M_{k-q_1}
M^*_{k-q_2} M_{k'-q_1} M^*_{k'-q_2}
\int_{-\infty}^{+\infty}d\tau \int_{-\infty}^{0}dt'
\int_{-\infty}^{0}dt  \nonumber  \\
&\times &\langle 0|
c^{\dagger}_{k'-q_1}(t'+\tau) c_{k'-q_2}(\tau)
c^{\dagger}_{k-q_1}(0) c_{k-q_2}(t)|0 \rangle
e^{-i(\omega_1+\epsilon_d)(t-t')-i(\omega_2-\omega_1)\tau}
\label{1}
\end{eqnarray}
Here $q_1$, $\omega_1$ ($q_2$, $\omega_2$) are the wave
vector and frequency of the incident (scattered) photons,
$c_k$ ($c^{\dagger}_k$) is the annihilation (creation)
operator of the conduction band electron, $\epsilon_d$
is the eigenenergy of the core hole, and $M_k$ is the optical
matrix element. The {\em
quasi-particle} approximation can be used to simplify
the calculation as well as to "pick-up" only the virtual
RS process contribution. In this approximation, we substitute
$c^{\dagger}_{k'-q_1}(t'+\tau)=e^{i\epsilon_{k'-q_1}t'}
c^{\dagger}_{k'-q_1}(\tau)$ and
$c_{k-q_1}(t)=e^{-i\epsilon_{k-q_1}t} c_{k-q}(0)$ \cite{mcw}
in Eq.[\ref{1}], and we obtain
\begin{eqnarray}
W(q_1,q_2;\omega_1,\omega_2) &=& \sum_{k,k'}
\frac{M_{k-q_1}M^*_{k-q_2}}{\omega_1-(\epsilon_{k-q_1}-\epsilon_d)}
\frac{M_{k'-q_1}M^*_{k'-q_2}}
{\omega_1-(\epsilon_{k'-q_1}-\epsilon_d)}
\nonumber  \\
&\times &\int_{-\infty}^{+\infty}d\tau
\langle 0|c^{\dagger}_{k-q_1}(\tau) c_{k-q_2}(\tau)
c^{\dagger}_{k'-q_1}(0) c_{k'-q_2}(0)|0 \rangle
e^{-i(\omega_2-\omega_1)\tau}
\label{2}
\end{eqnarray}
If we treat $M_q$ as a constant, and assume the RS is
near but not at resonance, {\em i.e.\/} incident
photon frequency $\omega_1 \sim \Delta_g$ but
$|\omega_1-(\epsilon_{k-q_1}-\epsilon_d)| >>\epsilon_{k-q_1}$,
where $\Delta_g$ is the semiconductor band gap,
then we obtain the celebrated form \cite{mcw,rs70}
$W(q_1,q_2;\omega_1,\omega_2) =
\frac{|M|^4}{(\omega_1-\Delta_g)^2} S(q_1-q_2,\omega_1-\omega_2)$
and the Random-Phase Approximation (RPA) can be used
to calculate the RS efficiency from collective excitations.
In this case, the RS efficiency is proportional to the dynamical
structure factor $S(q_1-q_2,\omega_1-\omega_2)$
with an enhancement prefactor.
Although this is a very useful result, its applicability in the
{\em resonance} region is invalid, especially for the 1D electron
gas.

Since the 1D TL model is exactly solvable even
including the core hole
effect, it would be of interest to study RS from the TL liquid.
Before including the core hole effect, we show that
taking into account the $k$-dependence of $M_k$
there is a contribution in Eq.[\ref{1}] which is proportional
to the structure factor $S(q_1-q_2,\omega_1-\omega_2)$ for
RS from the TL liquid. Expand $M_k$ around some point $k_0$
\cite{note2} and $\epsilon_k$ around some point $k_f'$ ($k_f'\sim$
Fermi momentum $k_f$). Since in resonance RS,
we can choose the appropriate $k_0$ and $k_f'$, then
$\frac{M_{k-q_1}M^*_{k-q_2}}
{\omega_1-(\epsilon_{k-q_1}-\epsilon_d)}
\sim \frac{|M|^2}{\omega_1-(\epsilon_{k-q_1}-\epsilon_d)}
-\frac{2Re(M'M^*)}{v_f}$.
Here $v_f$ is the Fermi velocity. It is
easy to see that the 2nd term will give a contribution
proportional to the structure
factor $S(q_1-q_2,\omega_1-\omega_2)$
in Eq[\ref{1}]. The relative
amplitude of this contribution depends on detailed
information about the system.

Let us now calculate the RS efficiency for the spinless
TL liquid with constant matrix elements $M_k=M$.
Experimentally the spinless model can be realized by
applying a strong magnetic field that polarizes electron
spins. Generalization to the spin-$1/2$ case is straightforward
but the calculation is more complicated.
In the standard bosonization \cite{1dgas}
of the 1D electron gas, the
Hamiltonian of the coupled TL bosons and localized
core hole is \cite{note1}:
\begin{equation}
H_0=\sum_k v_{TL}|k| b^{\dagger}_k b_k
+\sum_x \epsilon_d d^{\dagger}_x d_x +\frac{1}{\sqrt{L}}
\sum_{k,x} V_k e^{ikx} d^{\dagger}_x d_x b_k +h.c.
\label{4}
\end{equation}
The electron-radiation interaction can be written as:
\begin{equation}
H_{int}=M\sum_{x,q,\alpha} \psi^{\dagger}_{\alpha}(x) d_x
A_q e^{iqx} +h.c.
\label{5}
\end{equation}
where $\psi^{\dagger}_{\alpha} (x)$ is the fermion
operator of right moving ($\alpha=+$) and left moving
($\beta=-$) electrons,
and $b^{\dagger}_k$ and $b_k$ are the operators of the
TL boson whose velocity is
$v_{TL}=[(v_f+V_{1k})^2-(V_{2k})^2]^{1/2}$. $d_x$ and
$d^{\dagger}$ are operators of the localized core hole
with energy $\epsilon_d$.
Here we only include the forward scattering of the
electron by a core hole in Eq.[\ref{4}], the effects
of backscattering of electrons will be discussed in the end.
$V_k$ is the renormalized
interaction between the TL boson and the core hole
$V_k=V_{3k}(\frac{|k|}{2\pi})^{1/2}[\cosh\phi_k -\sinh\phi_k]$.
Here we use the notation as in \cite{mahan} for $V_{1k}$,
$V_{2k}$ and $\phi_k$. $V_{1k}$
and $V_{2k}$ are for the forward Fermion scattering interaction,
so in "$g$-ology", $V_{1k} \equiv g_4$ and
$V_{2k} \equiv g_2$.
$V_{3k}$ is the interaction
between electron and core hole. In the further calculation,
we will use the short-range $\delta$-potentials for
$V_i(x)$. The effect of $V_1$ is a shift of the Fermi
velocity, so we will define $\tilde{v}_f=v_f+V_1$. For the
$\delta$-potentials, the effective Fermion-Fermion coupling constant
$g \equiv \sinh^2(\phi_k) =
\frac{1}{2}[1/ \sqrt{1-v^2_2/ \tilde{v}^2_f} -1]$.
To calculate the transition rate $W(q_1,q_2;\omega_1,\omega_2)$, we
need to calculate an $8$-point correlation function  if we include the
core hole effect:
\begin{equation}
{\cal C}(x,x',u,t',t)=
\langle 0|
\psi_{\alpha}(x',t') d^{\dagger}_{x'}(t')
d_{x'}(u) \psi^{\dagger}_{\alpha}(x',u )
\psi_{\beta}(x,0) d^{\dagger}_x(0)
d_x(t) \psi^{\dagger}_{\beta}(x,t) |0 \rangle
\label{6}
\end{equation}
Using a unitary transform \cite{schotte},
the $8$-point correlation function decouples
to two $4$-point correlation functions, which
can be calculated easily. The transition rate is calculated
as:
\begin{eqnarray}
W^{\alpha \beta}(q_1,q_2;\omega_1,\omega_2) &=& const \int dy
e^{i v_{TL} (q_1-q_2)y} \int du \int dv
e^{-i (\omega _2 u+\omega _1 v)}
\nonumber  \\
&\times&\frac{f(u,v,y)}{[(u +y-i\epsilon)(v -y+i\epsilon)]^{
g_{\alpha \beta}^+} [(u-y-i\epsilon)(v+y+i\epsilon)]^{
g_{\alpha \beta}^-}}
\label{8}
\end{eqnarray}
With
\begin{equation}
f(u,v,y)= \int_{\frac{|u +v |}{2}}^{\infty} dx
\frac{[4x^2-(u-v+2y-2i\epsilon)^2]^{g_{\alpha \beta}^+}
[4x^2-(u-v-2y-2i\epsilon)^2]^{g_{\alpha \beta}^-}}
{[4x^2-(u+v+2i\epsilon)^2]^{\eta}}
\label{9}
\end{equation}

Here $\epsilon$ is the inverse of
the ultraviolet momentum cutoff,
and we have renormalized the photon frequencies by
$\omega_i \rightarrow \omega_i -(\epsilon_f-\bar{\epsilon_d})$.
The $g's$ are defined as:
$g_+=
g_{++}^+ =g_{--}^- =[\sqrt{g+1}-\delta \, (\sqrt{g+1}-\sqrt{g})]^2$;
$g_-=
g_{++}^- =g_{--}^+ =[\sqrt{g}-\delta \, (\sqrt{g+1}-\sqrt{g})]^2$;
$g_{\alpha \beta}^i =\sqrt{g_+ g_-}$
($\alpha \neq \beta$). A useful combination is
 $\eta=g_+ + g_-$. Here $\delta=V_3/v_{TL}$ is the phase shift
of {\em noninteracting} electrons scattered by $V_3$ \cite{schotte}.

Our exact results Eq.[\ref{8}-\ref{9}] are similar to that
in Ref.\cite{norabr}, where Nozi\`{e}res and Abrahams discussed
the threshold behavior of RS from {\em noninteracting}
electron systems. But they did not discuss the interference
effects. If we neglect the space-dependence and the
electron-electron interaction
effects, we will get the same results as in Ref.\cite{norabr}.
Note however the exponents we calculated here are different
because of the coupling of the core hole to both branches
($k$ and $-k$) of the electrons.

Now let us calculate the transition rate
$W(Q,\omega _1,\omega _2)$ with $Q=q_1-q_2$.
We drop the superscrpt $\alpha=\beta=+$ used in Eq.[\ref{8}]
and write it later. We can obtain
the result for $\alpha=\beta=-$ by
changing $Q \rightarrow -Q$ in $W(Q,\omega _1,\omega _2)$.
The calculation for
$\alpha \neq \beta$ is similar and we will only
discuss the results. It's easy to see there is a divergence in
Eq.[\ref{9}], and this divergence
can be cured by introducing  the core hole
lifetime $\tau$. In addition to removing the unphysical
divergence, we introduce the core hole lifetime $\tau$,
as in Ref.\cite{norabr} in order to be able to separate "hot
luminescence" (HL) and resonance Raman scattering (RRS)
in the total secondary emission given by Eq.[\ref{8}].
Following Ref.\cite{norabr},
$f(u,v,y)$ can be split into two parts:
$f(u,v,y)=\tau + \frac{1}{2} |u+v|\Psi(
\frac{u-v-2i\epsilon}{|u+v|}, \frac{y}{|u+v|})$ with

\begin{equation}
\Psi(z,y)=
\int_1^{\infty} dx \{[\frac{x^2-(z+y)^2}{x^2-1}]^{g_+}
[\frac{x^2-(z-y)^2}{x^2-1}]^{g_-} -1 \} -1
\label{10}
\end{equation}

After substituting Eq.[\ref{9}],
the first term in Eq.[\ref{8}] which is proportional
to $\tau$ will give the
{\em hot luminescence} (HL) contribution, and the second
term will give the {\em Raman Scattering} (RS)
contribution. Define $\omega_{\pm}=\omega_1-\omega_2
\pm Q v_{TL}$, for $ \eta >1 $, then the
HL contribution to the transition
rate is \cite{note4}:
\begin{eqnarray}
W_F(Q,\omega _1,\omega _2) = const \;
\Theta (\omega _1) \Theta (-\omega _2)
(-\omega _2)^{\eta -1} e^{-\epsilon (\omega _1 -\omega _2)}
\int _0^1 ds s^{g_- -1} (1-s)^{g_+ -1}
\nonumber \\
(2\omega_2s+\omega_-)^{g_- -1}
(2\omega _1 - 2\omega _2 s -
\omega_-)^{g_+ -1}
\Theta (2\omega _2 s +\omega_-)
\Theta(2\omega_1 - 2\omega_2 s -\omega_-)
\label{11}
\end{eqnarray}

The numerical calculation of $W_F$ is shown in
Fig.(1). The calculation shows that there is singular
behavior around $\omega _1-\omega _2 =|Qv_{TL}|$. Since
$\omega_2 <0$, so from $\Theta(2\omega_2 s+\omega_-)$
and $\Theta(2\omega_1-2\omega_2s-\omega_-)$ we have
$W_F(Q,\omega _1,\omega _2)
\propto \Theta (\omega _1-\omega _2 -|Qv_{TL}|)$. The
other two peaks are at $\omega _1+\omega _2 =\pm Qv_{TL}$.
The asymptotic behavior of $W_F$ can be given at these
singular peaks using the following integral
\begin{eqnarray}
\int^c_0 x^u (c-x)^v (x+a)^w =
c^{1+u+v}a^w {\Gamma(1+u)\Gamma(1+v) \over \Gamma(2+u+v)}
{ }_2F_1(-w,1+u,2+u+v,-c/a)
\label{eq:int1}
\end{eqnarray}
Where ${ }_2F_1(-w,1+u,2+u+v,-c/a)$ is the Hypergeometric
function. Without losing generality, we assume here
$Qv_{TL} \ge 0$.
Define $c=(\omega_1-\omega_2-Qv_{TL})/(-2\omega_2)$
and $\varepsilon=(\omega_1+\omega_2+Qv_{TL})/(-2\omega_2)$,
then $\varepsilon=c+({Qv_{TL} \over -\omega_2}-1)$. Now we discuss
various limits.
\newline
(i) at $c \stackrel{>}{\sim} 0$,
and $0 < \varepsilon <Qv_{TL}/(-\omega_2)$ [as in Fig.(1)]
\begin{eqnarray}
W_F &\sim &  (-\omega_2)^{-\eta-3} \int_0^cds
s^{g_--1} (c-s)^{g_--1}(s+\varepsilon)^{g_+-1}
\nonumber \cr
&\sim &  c^{2g_--1}\varepsilon^{g_+-1}{ }_2F_1(1-g_+,g_-,2g_-,-c/\varepsilon )
\label{eq:c1}
\end{eqnarray}
If $c \ll \varepsilon$, then $W_F \sim c^{2g_--1}$;
If $c \gg \varepsilon$, then
$W_F \sim C_1 c^{2g_--g_+}\varepsilon^{2(g_+-1)}
+ C_2 c^{3g_--1}\varepsilon^{g_+-1-g_-}$, where $C_i$ is constant.
If $c \sim \varepsilon$, then $W_F \sim c^{g_-+\eta-2}$.
\newline
(ii) at $c \stackrel{>}{\sim} 0$,
and $\varepsilon <0 $
\begin{eqnarray}
W_F &\sim & (-\omega_2)^{-\eta-3} \int_{-\varepsilon}^cds
s^{g_--1}(c-s)^{g_--1}(s+\varepsilon)^{g_+-1}
\nonumber \\
&\sim &
\Theta(c-|\varepsilon|)(c-|\varepsilon|)^{\eta-1}|\varepsilon|^{g_--1}
{ }_2F_1(1-g_-,g_+,2\eta,1-c/|\varepsilon|)
\label{eq:c2}
\end{eqnarray}
If $c \gg |\varepsilon|$, then
$W_F \sim C_1 c^{g_+}|\varepsilon|^{2(g_--1)}
+ C_2 c^{g_--2g_+-1}|\varepsilon|^{g_--1-g_+}$.
If $c \sim |\varepsilon|$, then
$W_F \sim (c-|\varepsilon|)^{\eta-1}|\varepsilon|^{g_-1}$.
\newline
(iii) at $\varepsilon \sim Qv_{TL}/(-\omega_2)$
then $c \sim 1$.
Define $\Delta = c-1$, then
\begin{eqnarray}
W_F &\sim& (-\omega_2)^{-\eta-3} \int_0^1ds
s^{g_+-1}(s+\Delta)^{g_--1}\Theta(s+\Delta)
\nonumber \\
&\sim& \cases{C_1\Delta^{2g_--2}+C_2 \Delta^{\eta-1},
& if $\Delta \ge 0$; \cr
C_1|\Delta|^{2g_+-2}+C_2 |\Delta|^{g_+-g_--1},
& if $\Delta < 0$ \cr}
\label{eq:c3}
\end{eqnarray}
(iv) In the above cases, we have assumed $\omega_2 \not\simeq 0$.
If $|\omega_2| \ll \omega_-$
and $|\omega_1 + \omega_2 \pm Qv_{TL}| \gg |\omega_2|$, then
\begin{equation}
W_F \sim (-\omega_2)^{\eta-1} (\omega_-)^{g_--1}
(\omega_1+\omega_2+Qv_{TL})^{g_+-1}
\label{eq:c4}
\end{equation}

For $\eta<1$, the HL contribution will diverge. We believe this
divergence comes from the interference effect
which we included.
The calculation of the RS part contribution for
$\eta<1$ is an extremely tedious task. However
since in this case the HL part will diverge and the
integral of the RS converges well, we will
not discuss the calculation in this region.
For $\eta >1$, the integration in Eq.[\ref{10}]
can be estimated around the lower limit of the integration.
The RS
part can be calculated:

\begin{eqnarray}
 W_R(Q,\omega _1,\omega _2) &=& const \;
(\omega_1 + \omega_2)^{\eta-1} \epsilon^{1-\eta}
\Theta (\omega _1-\omega _2)
[\delta (\omega_-) + O(\epsilon)
\tilde{f}_-(\omega_-)]
\nonumber  \\
& & [\delta (\omega_+) + O(\epsilon)
\tilde{f}_+(\omega_+)]
 + \cdots
\label{12}
\end{eqnarray}
Here $\tilde{f}_-(\omega_-)$ and $\tilde{f}_+(\omega_+)$
are smooth functions of $\omega_-$ and $\omega_+$ respectively.

We can see that there is a peak around
$\omega _1-\omega _2 =|Qv_{TL}|$.
The prefactor $(\omega _1+\omega _2)^{\eta-1}$
clearly shows that this peak has a different
origin from the peak of the structure factor.

In the first order optical process such as
photoemission and absorption, the competition
between electron-electron interaction and electron-hole
coupling in the TL is manifest in the FES exponent; the condition
for the critical point of positive and negative exponents
is $\eta=1$ \cite{fest1}. In secondary optical processes,
because of the interference of light from recombination
of holes and electrons at different sites, the spectrum
is more complicated, the competition of  electron-electron
interaction and electron-hole coupling will be manifest in
the RS part as well as HL part. But the critical point
is the same ($\eta=1$).

Now we turn to discuss $W^{\alpha \beta}$ with
$\alpha \neq \beta$, which is related to the processes
which describes the situation where
photon excited electron and the recombination
electron are in different branches. For $\eta >1$, the RS
contribution is the same as Eq.[\ref{12}]. The HL contribution
is proportional to $\tau^{1-(\sqrt{g_-}
-\sqrt{g_+})^2}<<\tau$. So
for $\sqrt{g_- g_+} >1/2$,  the HL contribution
can be neglected. If we assume the core hole lifetime
$\tau$ is larger than the characteristic time scale,
which will cure the divergence when $\eta <1$, the
HL contribution for $\sqrt{g_-g_+}<1/2$
is also negligible compared to the $\alpha
=\beta$ case.

It is tempting to discuss the threshold behavior of RS
from a higher-dimension (2D or 3D) electron
system as Schotte and Schotte
did for FES in emission (absorption) \cite{schotte}.
However, the possibility of extending our result
to 3D depends essentially on the scattering phase shift
being pure $s$-wave, which is not true in our
case, since we do include interference between different
core hole states. Technically this prevents
carrying through a partial wave expansion which is needed
for the 1D $\rightarrow$ 3D analogy to be valid from
TL liquid absorption/emission to threshold optical behavior
of the 3D Fermi gas.
But we can still use our result to discuss the interference
effects and interaction effects in Fermi liquid in high
dimensions. Following the recent arguments
in renormalization
group theory of Fermi liquid \cite{shankar},
this is especially true in
the threshold region of the secondary optical processes.
When neglecting the interference effects, similar
to the absorption (emission) case, the threshold behavior
of RS in a high dimension (2D or 3D) non-interacting
Fermion gas is similar to that of
the TL liquid \cite{norabr}. From our above results,
one effect of the interference which seems to be true
in a high dimensional Fermion gas
is that $W_F$ can {\em not} be factorized
into absorption and emission
any more. Similar effects have been found in RS from
polaritons \cite{ztb}. In Fig.[1], there are features
around $\omega_1-\omega_2=|Qv_{TL}|$ which suggest
$W_F$ looks more like Raman Scattering than luminescence.

A more striking effect in the TL liquid is that
the transition rate is very sensitive to the coupling strength
of TL bosons to the core hole, and the value of $g$.
We show in Fig.[2] the range of $V_2$ and $V_3$ that gives the
divergent and convergent HL $W_F$. We can see from Fig.[2]
that large $g$ (or $V_2$) and small $V_3$
is needed to have convergent $W_F$. A interesting result is
that the $W_F$ diverges for a 1D Fermi Liquid ($g=0$).
So the HL part of contribution will dominate the secondary
optical processes. The coupling of electron-hole
($V_3$) will increase this effect. When the electron-electron
interaction is strong enough, the HL part
$W_F$ will converge and
the RS $W_R$ will dominate the total secondary emission. Since
the HL part and RS part have different peak (singular)
structure, the theory does show different secondary optical
properties between TL and FL.

Strictly speaking, the HL part and RS part
can not be called "Hot Luminescence"
and "Raman Scattering" separately, since $W_F$
can not be factorized into absorption and
emission in the usual sense, and there is singular
structure around $\omega_1 -\omega_2 = |Q v_{TL}|$
in $W^F$. But we think of them as
having originated from "Hot
Luminescence" and "Raman Scattering" respectively
although the interference effects make them
non-separable.

Our main objective in this Letter has been to give an
exact model of secondary optical process in the resonance
region. For clarity and simplicity, we used a model of a
localized hole coupling with the TL without backscattering.
Also the electron-electron interaction is taken as
short-range ($\delta$-potential).
The effects of finite hole mass and unscreened
long range electron-electron interaction need further
study. The most important effect is that we omitted the backscattering
of the electron by core hole in TL liquid.
In the point potential, \cite{gogolin,kane-bs} each correlation function
in Eq.(\ref{8}) can be written as product
of a correlation function due to forward and backscattering
potential $G=G_f G_b$. So there are addtional contributions
to both $W_F$ and $W_R$.
It was shown
by a RG calculation in Ref.\cite{kane} that a
TL liquid with repulsive interaction
will be decoupled into two
separate TL liquids by even a weak impurity
if the  backscattering is taken into consideration.
There are recent studies of backscattering effects on the
FES in photoemission and photoabsorption \cite{gogolin,kane-bs,prokofev}.
In the weak
interaction limit, Kane {\it et al.} showed that
there is a crossover energy between strong and weak backscattering
for $\epsilon^* \sim (V_b/\hbar v_f)^{1/(1-g)}$. At $\epsilon <
\epsilon^*$, $G_b$ has power-law behavior, and the exponents
due to backscattering are positive.
We suggest this result is generally true even for
strong electron-electron interactions \cite{gogolin,prokofev}.
Since the divergence of the HL is due to the low
energy excitations, the backscattering effect will
make the HL converge better. If the core hole lifetime
is smaller than $\epsilon^*$, the HL contribution
$W_F$ will be changed mainly due to the increase of $g_{\pm}$.
If the core hole inverse lifetime is larger than the backscattering
energy interaction of electron and core
hole, the  backscattering effect on $W_F$ will be less dramatic, and
we suggest here the main feature of $W_F$ will not be
changed.
The RS is a virtual process and independent of the
core hole lifetime.
{}From Eq.(\ref{12}), we think the backscattering effect will only
change the prefactors of the resonance peak
at $\omega_1 - \omega_2 =|Qv_{TL}|$.

In conclusion, we have studied the RS from a TL liquid.
If we treat the electron-photon
matrix element as constant {\it i.e.}
independent of wave-vector, the
RS efficiency can not be simplified
to be proportional to the structure factor
$S(q_1-q_2,\omega_1-\omega_2)$. Generally it has
some complicated structure which is sensitive both to
the coupling between the electron and core
hole and the electron-electron interaction strength of the TL liquid.
If we take into account the $k$-dependence
of $M_k$,
in the resonance region, the RS efficiency does have some part of
the contribution which is proportional to
the structure factor $S(q_1-q_2,\omega_1-\omega_2)$.
We found the total secondary emission can not
be rigorously separated into {\em Hot Luminescence} and
{\em Raman Scattering}.

{\bf Acknowledgement}: We thank Prof. David Schmeltzer
for many stimulating discussions especially for the
effects of backscattering.
We also would like to thank Dr. A. M. Finkel'stein,
Dr. P. Littlewood,
Dr. P. M. Plaztman and Dr. A. Kuklov
for helpful discussions.
The work was supported
in part by FRAP-PSC-CUNY and by the NSF-INT-9122114.

\newpage
\begin{center}
\bf{FIGURE CAPTIONS}
\end{center}

\vspace{2cm}

{\bf Fig.1} Total HL transition rate $W_F^{++}+W_F^{--}$ as
a function of $\omega_1$ (incident photon frequency) and
$\omega_2$ (scattered photon frequency) at
$g_+=0.8$, $g_-=0.4$.
The unit for the $x$, $y$ axes is $|Q|v_{vf}$.
The positions
of the two peaks are at $\omega_1+\omega_2=\pm Qv_{TL}$.
The singular line is at $\omega_1-\omega_2=|Q|v_{TL}$.
Inset (a): The cross section at
$\omega_1-\omega_2=|Q|v_{TL}+0^+$; Inset (b): The cross section
at $\omega_1=\sqrt{2} |Q|v_{TL}$.

{\bf Fig.2} The divergent region and convergent region of
$W_F$.


\begin{references}
\bibitem{note2} Here $k_0 \sim k'_f+q$ is chosen to cancel the
denominator.
\bibitem{note1} We have dropped from the equations
of this paper all the operators (such
as $N$, $J$, $U_p$) which only
come into the chemical potential.
\end{references}
\end{document}